\documentclass{aa}

\usepackage{graphicx}
\usepackage{txfonts}
\begin{document}

\title{Propagating wave in active region-loops, located over the solar disk observed by
       the \emph{Interface Region Imaging Spectrograph}
       }

\author{B. Zhang \inst{1,2}, Y. J. Hou\inst{1,2}, \and J. Zhang\inst{1,2}
        }

\institute{CAS Key Laboratory of Solar Activity, National Astronomical Observatories,
           Chinese Academy of Sciences, Beijing 100012, China\\
           \email{yijunhou@nao.cas.cn; zjun@nao.cas.cn}
           \and
           School of Astronomy and Space Science, University of Chinese Academy of Sciences,
           Beijing 100049, China
           }

\date{Accepted 04 January 2018}


  \abstract
   {
   }
   {We aim to ascertain the physical parameters of a propagating wave over the solar disk
   detected by the \emph{Interface Region Imaging Spectrograph} (\emph{IRIS}).
   }
   {Using imaging data from the \emph{IRIS} and the \emph{Solar Dynamic Observatory} (\emph{SDO}), we
   tracked bright spots to determine the parameters of a propagating transverse wave in active
   region (AR) loops triggered by activation of a filament. Deriving the Doppler velocity of Si IV line
   from spectral observations of \emph{IRIS}, we have determined the rotating directions of active region
   loops which are relevant to the wave.
   }
   {On 2015 December 19, a filament was located on the polarity inversion line of the NOAA AR 12470. The
   filament was activated and then caused a C1.1 two-ribbon flare. Between the flare ribbons, two rotation
   motions of a set of bright loops were observed to appear in turn with opposite directions. Following
   the end of the second rotation, a propagating wave and an associated transverse oscillation were detected
   in these bright loops. In 1400 {\AA} channel, there was bright material flowing along the loops in a
   wave-like manner, with a period of $\sim$128 s and a mean amplitude of $\sim$880 km. For the transverse
   oscillation, we tracked a given loop and determine the transverse positions of the tracking loop in a
   limited longitudinal range. In both of 1400 {\AA} and 171 {\AA} channels, approximately four periods
   are distinguished during the transverse oscillation. The mean period of the oscillation is estimated as
   $\sim$143 s and the displacement amplitude as between $\sim$1370 km and $\sim$690 km. We interpret
   these oscillations as a propagating kink wave and obtain its speed of $\sim$1400 km s$^{-1}$.
   }
   {Our observations reveal that a flare associated with filament activation could trigger a kink
   propagating wave in active region loops over the solar disk.
   }

\keywords{Sun: atmosphere --- Sun: filaments, prominences --- Sun: flares --- Sun: oscillations}

\titlerunning{Propagating wave}
\authorrunning{Zhang et al.}

\maketitle
%

\section{Introduction}
The solar corona consists of ionized plasma with a temperature of millions of kelvins, about
three orders of magnitude higher than the temperature of the visible solar surface (Erd{\'e}lyi \&
Fedun 2007). The heating process that generates and sustains the hot corona is still an unresolved
problem in solar physics. It is believed that the subphotospheric convection zone supplies enough
kinetic energy to the corona, where sufficient energy may be dissipated resulting in the
heating of coronal plasma. Parker (1988) put forward that the magnetic reconnection of neighboring
stressed field lines with opposite polarity vector components would release magnetic energy violently
and contributes to heating the corona. This idea was followed by a large number of studies (Solanki
et al. 2003; Klimchuk 2006; Bradshaw \& Klimchuk 2015; Wright et al. 2017). A possible scenario of
energy transporting from the convection zone to the corona is that the convection below the solar
visible surface and solar global oscillations may produce magnetohydrodynamic (MHD) waves in the
photosphere, which then propagate upward into the corona carrying energy (Alfv{\'e}n 1947; Parker
1958; De Pontieu et al. 2007; Tomczyk et al. 2007; Antolin et al. 2017). In uniform plasmas, there
are three basic types of MHD waves: slow and fast magnetoacoustic waves, and Alfv{\'e}n waves.
According to the frequencies observed by solar instruments, MHD modes in cylindrical plasma structures
could be mainly grouped into four categories: kink, sausage, longitudinal, and torsional (Van Doorsselaere
et al. 2008a). The first three modes are magnetoacoustic and hence compressible while the last one is
the incompressible Alfv{\'e}n mode.

With the progress in the spatial and temporary resolution of solar observation instruments, abundant
evidence of various types of MHD waves (standing and propagating waves) have been detected in various
solar activities. Propagating and standing transverse waves have been observed in spicules (He et al.
2009; Zaqarashvili \& Erd{\'e}lyi 2009; Okamoto \& De Pontieu 2011; Pereira et al. 2016), in active
region (AR) fibrils (Pietarila et al. 2011; Jafarzadeh et al. 2017), in mottles (Kuridze et al. 2012,
2013), in filament threads (Lin et al. 2007; Terradas et al. 2008; Li \& Zhang 2012; Shen et al. 2014),
and in coronal loops (Ofman \& Wang 2008; Tian et al. 2012; Guo et al. 2015; Li et al. 2017). Most of
these events could be explained as the magnetic flux tubes being perturbed by an external excitation,
and this perturbation appears as transverse oscillation and propagates along the flux tube, which is
usually detected as a temporal variation of the magnetic tube position or the Doppler velocity
(Oliver et al. 2014).

\begin{figure*}
\centering
\includegraphics [width=1.\textwidth]{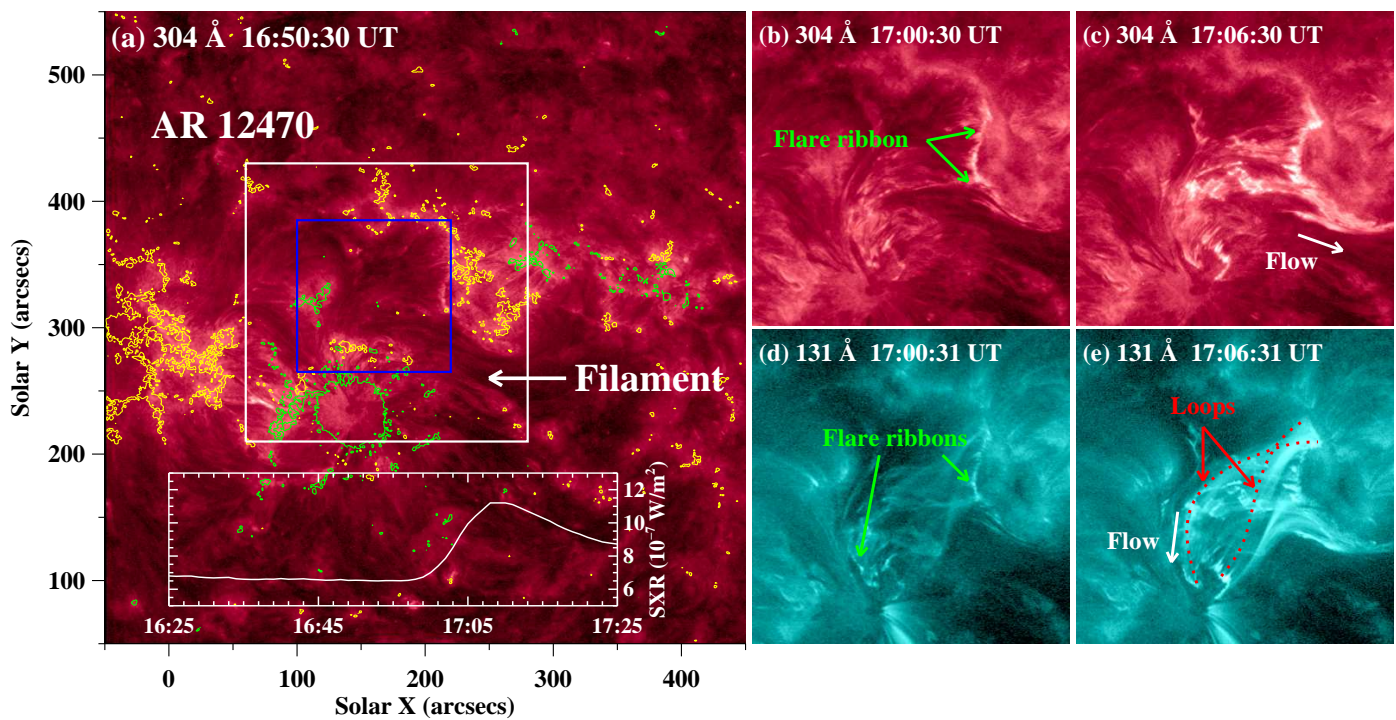}
\caption{
Panel (a): \emph{SDO}/AIA 304 {\AA} image displaying the overview of the NOAA AR 12470 on 2015 December 19.
The filament before activation is denoted by the white arrow. The white square outlines the field of view
(FOV) of panels (b)-(e) while the blue square represents the FOV of 1400 {\AA} images in Figure 2. The
green and yellow curves are contours of corresponding negative and positive magnetic fields, respectively.
The white curve in the bottom shows GOES SXR 1–8 {\AA} flux of the associated C1.1 flare.
Panels (b)-(e): sequence of AIA 304 {\AA} and 131 {\AA} images showing the flare caused by the filament's
activation in two different temperatures. The green arrows in panels (b) and (d) mark the flare ribbons
while the white arrows in panels (c) and (e) approximate the trajectory of the filament material
flows. The red dotted lines in panel (e) indicate the flare loops with shear in high temperature.
An animation (1.mpg) of the 131 {\AA} and 304 {\AA} images is available online.
}
\label{fig1}
\end{figure*}

In recent years, the new-generation satellites such as \emph{Solar Dynamic Observatory} (\emph{SDO};
Pesnell et al. 2012) and \emph{Interface Region Imaging Spectrograph} (\emph{IRIS}; De Pontieu et al.
2014) have provided observations that have high tempo-spatial resolution ranging from the photosphere,
chromosphere, and transition region to corona. Although there have been many studies of MHD
waves in the last few decades, observations about the propagating wave taken by the \emph{IRIS}
are rarely reported (Okamoto et al. 2015). Over the solar disk, there are only a few published
results of chromospheric waves from the \emph{IRIS} data (Bryans et al. 2016; Kanoh et al. 2016).
One important reason is that there are so many fine structures in the chromosphere, thus making it
difficult to detect the wave signal in a particular chromospheric structure among the abundant features.
By checking two years of \emph{IRIS} data (from June 2014 to May 2016), we find an example of a
wave-oscillation event over the disk on 2015 December 19. This event enables us to examine the
photospheric magnetic field environment. In present paper, we report on the wave-oscillation event
in active region loops triggered by filament activation which results in the storing and releasing
of twist in the loops.

The remainder of this paper is structured as follows. Section 2 contains the observations and data
analysis taken in our study. The observations of the filament activation, flare eruption, rotation
motions, and the subsequent propagating transverse wave in the loops are presented in Section 3.
Finally, in Section 4 we conclude this work and discuss the results.

\begin{figure*}
\centering
\includegraphics [width=1.\textwidth]{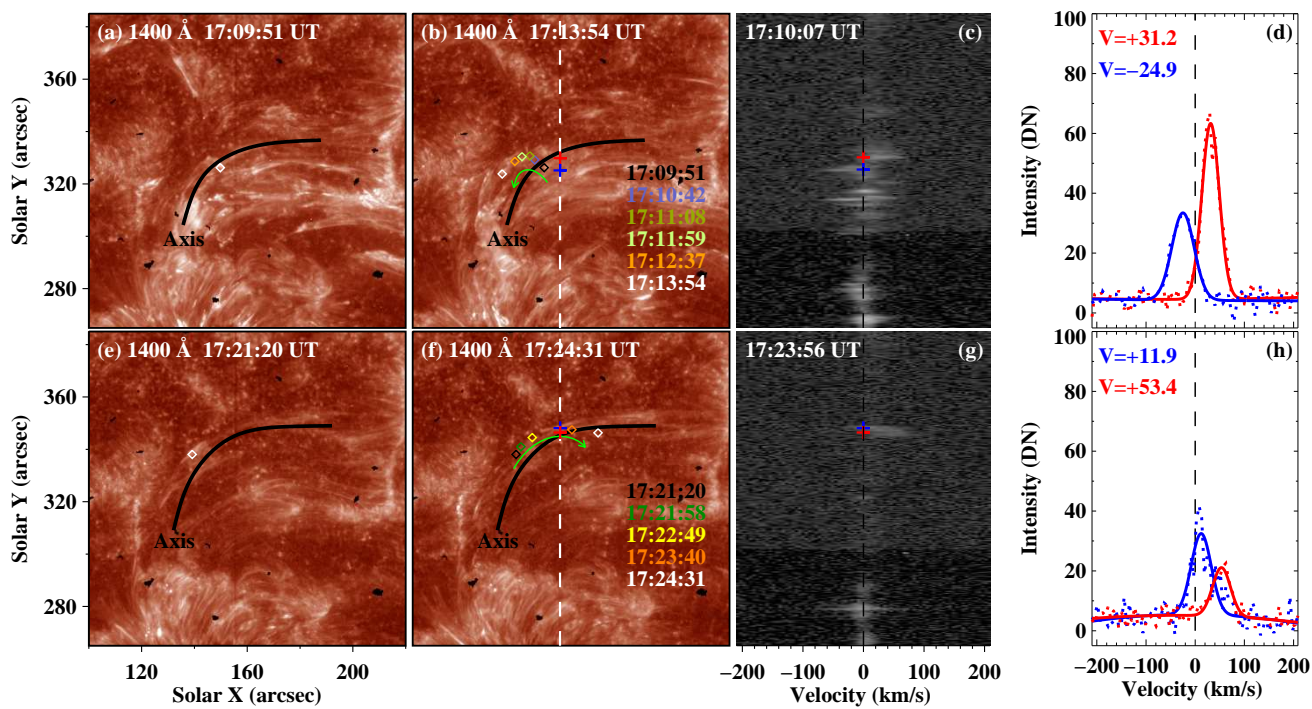}
\caption{
Panels (a)-(b): \emph{IRIS} SJIs of 1400 {\AA} displaying the first rotation motion around 17:10 UT
on 2015 December 19. The black line approximates the axis of the rotation while the green arrow marks
the direction of the rotation. The positions of the tracked feature in different times are denoted by
diamonds of different colors, of which the white one represents the current position. The white dashed
line in panel (b) shows the location of slit.
Panels (c)-(d): Si IV 1402.77 {\AA} line spectra along the slit in panel (b) and the profiles (dotted
lines) and its Gaussian fittings (solid lines) of this line at blue and red plus positions in panels
(b) and (c). The observation time is at 17:10:07 UT.
Panels (e)-(f): \emph{IRIS} 1400 {\AA} images exhibiting the second rotation motion of the loops just
before the appearance of a propagating wave. The features marked here are similar to those in panels
(a)-(b).
Panels (g)-(h): similar to panels (c)-(d), but at 17:23:56 UT.
The temporal evolution of the two rotation motions is available as an animation (2.mpg) online.
}
\label{fig2}
\end{figure*}

\section{Observations and data analysis}
On 2015 December 19, the \emph{IRIS} was pointed at NOAA AR 12470 from 16:41:13 UT to 17:37:33 UT,
with the central coordinate of (152{\arcsec}, 302{\arcsec}). We hence obtained a series of slit-jaw
1400 {\AA} images (SJIs) with a cadence of 13 s, a pixel scale of 0.{\arcsec}333, and a field of
view (FOV) of 167{\arcsec} $\times$ 175{\arcsec}. The 1400 {\AA} channel contains emission from
the Si IV 1394/1403 {\AA} lines that are formed in the lower transition region, which include the
contributions from continuum that is formed in lower chromosphere as well. The employed data are
all level 2, which have applied dark current subtraction, flat field, geometrical, and orbital
variation corrections (De Pontieu et al. 2014). For the spectroscopic analysis, we mainly adopted
the Si IV 1402.77 {\AA} line which is formed in the middle transition region with a temperature
of about 10$^{4.9}$ K (Li et al. 2014; Tian et al. 2014). The spectral data were taken in a large
sit-and-stare mode with a step cadence of 3.2 s. As the Si IV profiles were relatively close to a
single Gaussian, we used single-Gaussian fits to approximate the 1402.77 {\AA} line (Peter et al. 2014).

The Atmospheric Imaging Assembly (AIA; Lemen et al. 2012) and the Helioseismic and Magnetic Imager
(HMI; Schou et al. 2012) observations from the \emph{SDO} were also used. The AIA continuously
observes the multilayered solar atmosphere, including the photosphere, chromosphere, transition
region, and corona, in ten (E)UV passbands with a cadence of 12 s and a spatial resolution of
1.{\arcsec}2. The observations of 304 {\AA}, 171 {\AA}, and 131 {\AA} on 2015 December 19 were
employed in the present work. We also applied the corresponding full-disk line of sight (LOS)
magnetograms from the HMI, with a cadence of 45 s and a pixel size of 0.{\arcsec}5. Using the
standard routine ``hmi\_prep'' in Solar Software package, we calibrated the AIA and HMI data
to a specified pointing, rescaled the images to a spatial resolution of 1.{\arcsec}2, and
de-rotated these images (removing the roll angle) so that the solar east-west and north-south
axes were aligned with the horizontal and vertical axes of the image, respectively. Then these
data were all de-rotated differently to a reference time (16:41:13 UT, the start time of the
\emph{IRIS} observations). The \emph{IRIS} observations were also all de-rotated to the same
time. Finally, we aligned the \emph{IRIS} data with the \emph{SDO} observations according to
specific features such as bright spots and loops which were observed by both the \emph{IRIS}
and \emph{SDO}.

\section{Results}
\subsection{Activation of a filament and onset of the subsequent flare}

On 2015 December 19, NOAA AR 12470 was located near the solar disk-center. Around the main sunspot
of this AR, a C1.1 flare with two ribbons took place near 17:00 UT, which was caused by a filament's
activation (see animation 1.mpg). Plenty of filament material was brightened and then displayed
different dynamic evolutions. Partial bright material moved along the filament’s axis and slipped away.
However, some other bright material interacted with the active region loops which overlaid the filament.
These loops were disturbed by the filament material and underwent twice rotating motions from 17:09 UT
to 17:26 UT. Subsequently, a propagating wave and the associated transverse oscillation were detected.
Figure 1 shows the overview of the area of interest, the process of the filament's activation, and
the flare's onset in AIA 304 {\AA} and 131 {\AA} channels. The AR's main sunspot with negative
polarity was surrounded by positive fields. And a filament lied on the polarity inversion line (PIL)
of this AR (see panel (a)). The C1.1 flare started at 16:58 UT, and at the onset of the flare, the
brightening of the western footpoints of the flare loops gradually propagated toward the southwest
(see panel (b)), which developed into the western flare ribbon above positive magnetic fields. The
eastern flare ribbon of this flare was located in the sunspot umbra with strong negative field (panel
(d)). Between these two ribbons, the flare loops with evident shear appeared in 131 {\AA} channel (see
panel (e)), which is dominated by the emission from Fe XXI line [log T $\sim$7.05] (O'Dwyer et al. 2010).
Meanwhile, the activated filament material was detected to flow from the north-east to the south-west
with a projected speed of $\sim$90 km s$^{-1}$ (see panel (c)). Another flow was observed along the
flare loops, which fell down in the sunspot umbra (see panel (e)). This flare increased to its maximum
at 17:10 UT and faded away in the following ten minutes (see the \emph{GOES} flux curve in the bottom
of panel (a)).

\subsection{Rotations of active region loops}

Following the flare loops' appearance, two rotation motions were observed in the loops connecting
negative-polarity sunspot with northwestern positive fields (see animation 2.mpg). Around 17:09 UT,
the peak of the flare, a set of bright loops appeared in 1400 {\AA} images (see Figure 2(a)). Then the
first rotation motion took place during these loops, lasting from 17:09 UT to 17:15 UT. Examining the
AIA 304 {\AA} and 131 {\AA} observations (see animation 1.mpg), we noticed that some bright material
which came from the activated filament moved in the rotating loops. We chose one bright spot which is
relatively isolated in 1400 {\AA} channel and determine the positions of the tracked spot at different
times (see panel (b)). Looking from left to right along the rotation axis, the bright spot rotated
about the axis anticlockwise. To confirm this rotation motion, we employ the \emph{IRIS} spectra data
for Doppler velocity measurement. Panel (c) shows the Si IV 1402.77 {\AA} line spectra along the slit.
The positions at which the rotating loops cross the slit are indicated by red and blue plus symbols
in the figure. The spectra profiles and Gaussian fits of these two positions are shown in panel (d).
Here we get the Gaussian fit to the Si IV profiles by computing a non-linear least-squares fit to
function f(x) with six parameters. The function f(x) is a linear combination of a Gaussian and a quadratic:
$f(x)=A_{0}e^{\frac{-z^{2}}{2}} + A_{3} + A_{4}x + A_{5}x^{2}$, where $z=\frac{x-A_{1}}{A_{2}}$. The
six parameters represent the height of the Gaussian ($A_{0}$), the center of the Gaussian ($A_{1}$),
the width of the Gaussian ($A_{2}$), the constant term ($A_{3}$), the linear term ($A_{4}$), and the
quadratic term ($A_{5}$), respectively. Meanwhile, one-sigma error estimates of the six returned
parameters are also given. As a result, we calculate the uncertainty of the Doppler velocity according
to the one-sigma error estimate of the center of the Gaussian ($A_{1}$) (Li et al. 2016). At 17:10:07 UT,
the Doppler velocities at blue and red plus positions are respectively --24.9 $\pm$ 0.9 km s$^{-1}$
(blueshift) and 31.2 $\pm$ 0.5 km s$^{-1}$ (redshift). The pair of blueshift and redshift signals
provide a strong evidence for the rotation motion. Undergoing this rotation, a set of loops moved
northward and reached the northernmost point around 17:21 UT. Then, from 17:21 UT to 17:24 UT, the
second rotation motion of these loops was detected again. But this time, the rotation was clockwise
if one looks from left to right along the rotation shaft (see panel (f)). The spectral data at 17:23:56
UT were taken to investigate the second rotation motion. Through similar analysis method mentioned above,
we estimate the Doppler velocities at the positions indicated in the figure as 11.9 $\pm$ 1.8 km s$^{-1}$
and 53.4 $\pm$ 1.5 km s$^{-1}$ (panel (h)), respectively.

\subsection{A propagating wave with transverse oscillation in the loops}

\begin{figure}
\centering
\includegraphics [width=0.48\textwidth]{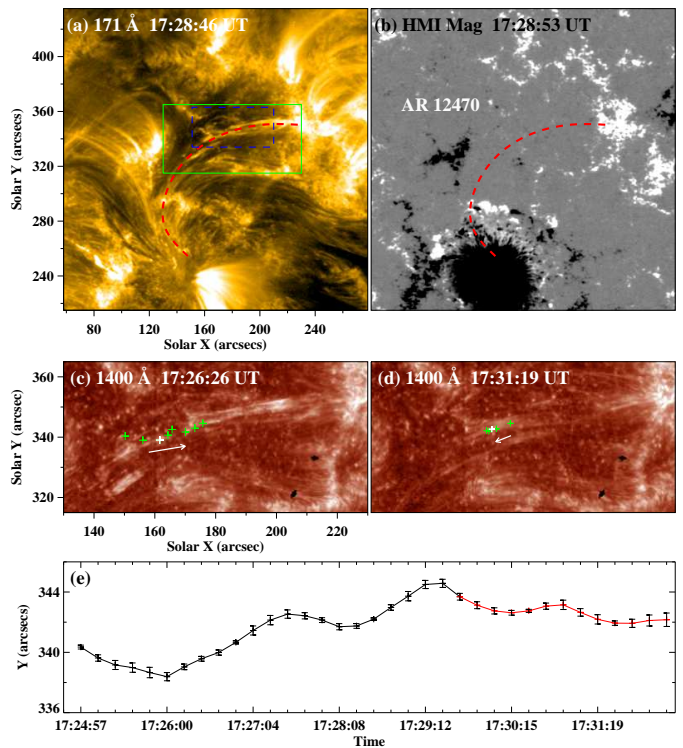}
\caption{
Panels (a)-(b): \emph{SDO}/AIA 171 {\AA} image and \emph{SDO}/HMI LOS magnetogram showing the EUV
loops of the focused region on 2015 December 19 and its underneath magnetic fields. The red dashed
lines delineate the magnetic loops along which the wave propagates. The green solid rectangle in
panel (a) outlines the FOV of panels (c)-(d) while the blue dashed rectangle represents the FOV
of 1400 {\AA} and 171 {\AA} images in Figures 4 and 5.
Panels (c)-(d): \emph{IRIS} SJIs of 1400 {\AA} displaying the propagation of the wave. The white
arrows denote the propagating directions of the wave by tracking a bright spot. The green pluses
outline the trajectory of the bright spot from 17:24:57 UT to 17:32:23 UT and the white pluses mask
the positions of the spot at the moments shown in panels (c) and (d). Panel (e): the evolution of
the spot's transverse position during the propagation of the wave. The black curve shows the wave
propagating to the right (panel (c)) while the red to the left (panel (d)).
An animation (3.mpg) of the 1400 {\AA} images is available online.
}
\label{fig3}
\end{figure}

The loops undergoing two rotation motions were clearly observed in 171 {\AA} passband too (see Figure
3(a)). Comparing the 171 {\AA} images with HMI magnetograms, we can determine that the loops originated
from the main sunspot with negative polarity and connected to the northwestern region with positive
fields (see panel (b)). At the later phase of the second rotation, several bright spots moved in a
wave-like manner from left to right along the active region loops. We chose one distinct bright spot
(with the maximum brightness within the target region) at 17:24:57 UT in 1400 {\AA} channel. We tracked
this spot and determined the coordinates of the spot frame-by-frame from 17:24:57 UT to 17:32:23 UT
(see animation 3.mpg). After repeating the tracking process for ten times, we obtained ten groups of
the spot's coordinates at each frame. Then, we got the average values of the spot's coordinates at
each frame and took the standard deviation as the uncertainty. At about 17:25 UT, this bright spot
moved along these loops toward the west with a mean projected velocity of 66.2 $\pm$ 1.1 km s$^{-1}$
(see panel (c)) and moved back to the east after 17:29 UT with an average projected velocity of 12.7
$\pm$ 1.3 km s$^{-1}$ (see panel (d)). The positions of the tracked feature at different times are
indicated in the figure. Panel (e) shows the bright spot's mean transverse positions with error bars
during the propagation of the wave. We can see that the position curve reveals a periodic oscillation,
with an average period of $\sim$128 s and amplitude of 880 $\pm$ 170 km.

\begin{figure}
\centering
\includegraphics [width=0.48\textwidth]{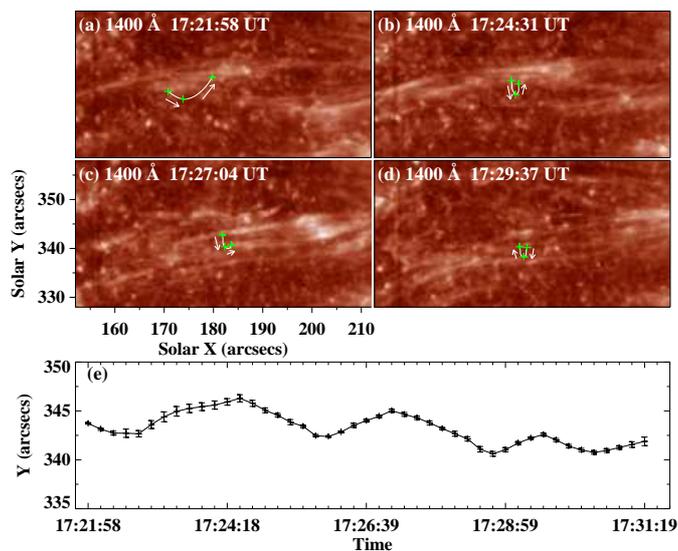}
\caption{
Panels (a)-(d): sequence of \emph{IRIS} 1400 {\AA} images showing the development of the transverse
oscillation from 17:21:58 UT to 17:31:19 UT. The green pluses and white lines approximate the track
of the loop's oscillation, and the white arrows indicate the moving directions of the tracked material.
The FOV of these panels has been outlined by the blue dashed rectangle in Figure 3(a).
Panel (e): the evolution of the tracking loop's transverse position during the oscillation.
The full temporal evolution of the oscillation in 1400 {\AA} is available as an animation (4.mpg) online.
}
\label{fig4}
\end{figure}

\begin{figure}
\centering
\includegraphics [width=0.48\textwidth]{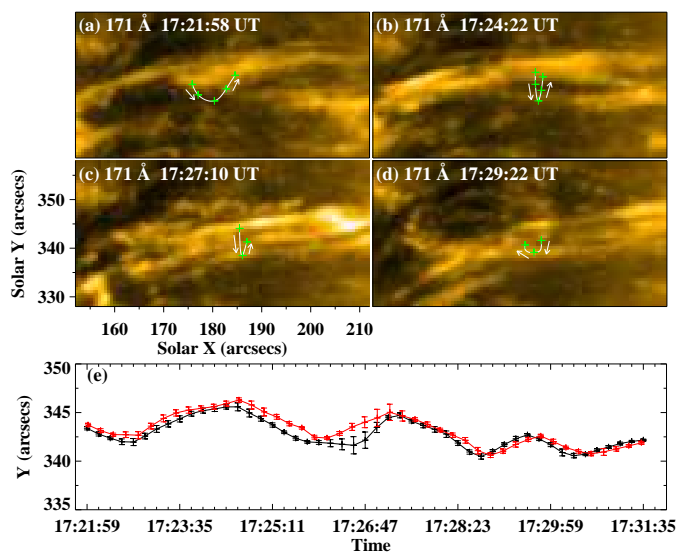}
\caption{
Similar to Figure 4, but in a higher temperature of the 171 {\AA} channel. The red curve are the
duplications of Figure 4(e).
}
\label{fig5}
\end{figure}

Also accompanying the second rotation, several bright loops tracked by the bright material display
regular transverse oscillation. To investigate the transverse oscillation of these loops, we tracked
a given loop from 17:21:58 UT to 17:31:19 UT. A bright spot inside this oscillating loop was tracked
in a limited longitudinal range (15{\arcsec}). We determined the coordinates of this spot at each
frame in 1400 {\AA} passband with the same method used in Figure 3. We then got the average values
of the spot coordinates at different frames and their standard deviations. Figure 4 shows the mean
transverse oscillation in 1400 {\AA} images (see animation 4.mpg). The transverse position trajectory
of the tracking loop is outlined in panels (a)-(d), and panel (e) shows the change of the oscillating
loop's transverse position versus time with the error bars. It is shown that the transverse oscillation
is visible for approximately four periods, with obvious damping - the amplitude decreases from 1370
$\pm$ 210 km to 690 $\pm$ 140 km. We counted the oscillation period as $\sim$143 s, mean upward velocity
as 28 $\pm$ 5 km s$^{-1}$, and mean downward velocity as 32 $\pm$ 4 km s$^{-1}$. Under the reminder of
the \emph{IRIS} observations, we noticed that the oscillation motion could be detected in a higher
temperature. We applied the 171 {\AA} data (see Figure 5) and analyzed them in the same way as for the
1400 {\AA} channel. The oscillation is visible also for four periods, and the oscillation parameters,
such as period, amplitude, and velocity, are nearly the same as those in the 1400 {\AA} bandpass (see
Figure 5(e)). We fitted the temporal evolution of the oscillation with a damped sine function described
in Nakariakov et al. (1999) and obtained the decay time of $\sim$500 s.

\section{Conclusions and discussion}
Employing the high-resolution observations from the \emph{IRIS} and the \emph{SDO}, we report a
propagating wave in active region loops, which is triggered by filament activation. On 2015 December 19,
a filament lay on the PIL of the NOAA AR 12470, which was activated and caused a C1.1 flare. Between
the flare ribbons, a set of bright loops were detected to first rotate anticlockwise and then rotate
clockwise. Following the end of the clockwise rotation, a propagating wave and an associated transverse
oscillation were detected in the loops. In the 1400 {\AA} channel, there was bright material moving
toward the west along the loops with a mean velocity of 66.2 $\pm$ 1.1 km s$^{-1}$ at first. Then,
some bright material moved back to the east with a mean velocity of 12.7 $\pm$ 1.3 km s$^{-1}$. To
investigate the transverse oscillation of these loops, we tracked a given loop and determined the
transverse positions of the tracking loop in a limited longitudinal range (15{\arcsec}). In both of
1400 {\AA} and 171 {\AA} channels, about four periods are distinguished during the transverse oscillation.
The period of the oscillation is estimated as $\sim$143 s, but the amplitude decreases from 1370
$\pm$ 210 km to 690 $\pm$ 140 km.

In the chromosphere, only a few examples have been reported of waves or oscillations detected in fine
structures over the disk. High resolution H$\alpha$ filtergrams obtained with the Swedish Solar Telescope
resolved evidence for small amplitude (1 – 2 km s$^{-1}$) waves propagating along a number of on-disk
filament threads. The oscillatory period of individual threads vary from three to nine minutes
(Lin et al. 2007). Observations of Ca II 8542 {\AA} revealed a kink wave in an on-disk chromospheric
active region fibril, and the properties of the wave were similar to those observed in off-limb spicules
(Pietarila et al. 2011). Kanoh et al. (2016) employed simultaneous \emph{Hinode} (Kosugi et al.2007)
and \emph{IRIS} observations of a sunspot umbra to derive the upward energy flux at photosphere and
lower transition region and estimate the energy dissipation. Temporal evolutions of the Doppler
velocity, the magnetic flux density, and the line core intensity derived from \emph{Hinode}/SP data
were applied to determine the periodic oscillating features at photosphere. Moreover, the Doppler
velocities derived from \emph{IRIS} Mg II K and Si IV lines were used to detect the oscillations at
the chromosphere and the lower transition region. They concluded that standing
slow-mode waves are dominant at the photosphere, and the high-frequency leakage of the slow-mode waves
is observed as upward waves at the chromosphere and the lower transition region. Bryans et al. (2016)
reported strong recurring upward propagating flows observed by \emph{SOD}/AIA with apparent speeds of
100 - 120 km s$^{-1}$ at coronal loop footpoints. Furthermore, magnetoacoustic shock waves were detected
by \emph{IRIS} Mg IIh line. These observations illustrated that both shock waves and flows are involved
at the footpoints. Different from previous works, here we first report a propagating wave with an
associated transverse oscillation in active region loops, which is triggered by filament activation.
In \emph{SDO}/AIA 171 {\AA} and \emph{IRIS} 1400 {\AA} channels, there is bright material
moving along the loops with a wave-like shape. Meanwhile, the loops display transverse oscillation.

Before the detection of the propagating wave, we observed two rotation motions in the bright loops.
The first rotation appeared between the two flare ribbons just at the peak of the flare around 17:10 UT.
In the 1400 {\AA} observation, the bright material in the rotating loops is rotated anticlockwise if one
looks from left to right along the rotation axis. To check this anticlockwise rotation, we analyzed the
Si IV 1402.77 {\AA} line spectra along the slit which crosses the rotating loops. The spectral profiles in
positions where the rotating loops cross the slit are approximated by single-Gaussian fits. Here we can
imagine that there existed a tube, and the bright loops wound round this tube. Since the Doppler
velocity in the lower position (blue plus) of the slit is --24.9 $\pm$ 0.9 km s$^{-1}$ (blueshift)
while that in upper position (red plus) is 31.2 $\pm$ 0.5 km s$^{-1}$ (redshift), we suggest that the
red plus is located in upper side of the tube while the blue plus is located in the lower side and
this rotation is anticlockwise when looking from left to right. Subsequently, the loops moved northward
and reached the northernmost point around 17:21 UT. They then performed the second rotation motion -
clockwise when looking form left to right along the rotation axis. Similar to the first rotation, we
chose two positions along the slit for spectral analysis. The Doppler velocity in the upper position
of the slit is 11.9 $\pm$ 1.8 km s$^{-1}$ (redshift) and that in lower position is 53.4 $\pm$ 1.5
km s$^{-1}$ (redshift). The same signal of Doppler velocity means that these two positions are located
in the same side of this tube. Combing the fact that the redshift velocity of the lower position is
bigger than that in the upper position, we propose that both of the two positions are located in the
lower side of the tube and this rotation is clockwise when looking from left to right. Since the two
rotation motions of these loops were in opposite directions, combining the fact that the first rotation
appeared just when the flare peaked and the second rotation resulted in the appearance of a propagating
wave, we propose that the energy released by the flare twisted the bright loops and was stored as
non-potential magnetic energy in the twisted loops during the first rotation motion. Then in the
second rotation, the stored energy was released by the unwinding motion of the twisted magnetic loops
(Yan et al. 2014), leading to a transverse wave propagating along these loops.

Waves and oscillations can be detected either by imaging or spectroscopic observations in solar limb
(Papushev \& Salakhutdinov 1994; Zaqarashvili et al. 2007; De Pontieu et al. 2007; He et al. 2009).
On the solar disk, quasi-periodic propagating disturbances have been found in large coronal loops
(e.g., Nakariakov et al. 1999; Nightingale et al. 1999; Berghmans \& Clette 1999; Aschwanden 2009;
Ruderman \& Erd{\'e}lyi 2009). The propagation of kink waves may lead to the observed periodic
transverse displacement of the loop axis. When the velocity of the kink wave is polarized in the plane
of observation, the spectral observation shows obvious Doppler shift in the location of oscillating
loop. But if the velocity is polarized in the perpendicular plane, the visible displacement of the
loop could be detected by the imaging observation. Since the observations in our study display
propagating displacements of the active region loops in the 1400 {\AA} and 171 {\AA} channels, we
interpret these oscillations as a propagating kink wave. This explanation is consistent with the
reports of low amplitude kink oscillations in the solar corona (Tomczyk et al. 2007; Van Doorsselaere
et al. 2008b). To estimate the density of the magnetic field in a fundamental kink oscillating loop,
we applied the formula $B = \frac{2L}{P}  \sqrt{ \mu m_{p} m_{m} (n_{in}+n_{ex})/2}$ (Roberts et al. 1984;
Aschwanden \& Schrijver 2011). The P is the period of oscillation, L is the length of the loop, $\mu$
is the magnetic permeability, $m_{m}$ is the molecular weight, $m_{p}$ is the proton mass, $n_{in}$
and $n_{ex}$ are the number densities of the proton inside and outside of the loop. We estimate that
L $\sim$ 100 Mm, P $\sim$ 143 s, $\mu$=4$\pi$, $m_{m}$=1.2 (Verwichte et al. 2013),
$m_{p}$ $\sim$ 1.673$\times$10$^{-24}$ g, ($n_{in}$+$n_{ex}$)/2 $\sim$ 9$\times$10$^{8}$ cm$^{-3}$
(Van Doorsselaere et al. 2008b). We then measured the magnitude of the magnetic field in the
oscillating loop to be $\sim$21 G.

The transverse oscillation presented in this work was observed with a decay time of about 500 s,
comparable to previous results (Nakariakov et al. 1999; Wang \& Solanki 2004;
White \& Verwichte 2012). The decay of the loop oscillation amplitude indicates the dissipation of
the wave energy, which contributes to the heating of the loops. To estimate the energy flux of
the wave, we applied the formula $E_{w}=\rho (\nu)^2 V_{A}$, where $\rho$ is the mass density, $\nu$ is
the velocity amplitude, and $V_{A}$ is the Alfv\'{e}n speed (Tomczyk et al. 2007). The mass density was
estimated to be $\rho = m_{p} m_{m} (n_{in}+n_{ex})/2$ $\sim$ 1.8$\times$10$^{-15}$ g cm$^{-3}$
(Van Doorsselaere et al. 2008b; Verwichte et al. 2013). We measured that the maximum displacement amplitude
($A_{max}$) is $\sim$1370 km and the oscillation period (P) has a mean value of $\sim$143 s.
Then the maximum velocity amplitude ($\nu_{max}$) is estimated to be $\sim$60 km s$^{-1}$ by the formula
$\nu_{max} = 2 \pi A_{max}/P$ (Ofman \& Wang 2008). Assuming the oscillation is a fundamental kink mode,
we obtained the kink speed $C_{k}$=2 L/P $\sim$ 1400 km s$^{-1}$ and the Alfv\'{e}n speed is
$V_{A}$=$C_{k}$ / $\sqrt{2}$ $\sim$ 990 km s$^{-1}$. Then we obtained the energy flux of the wave
$E_{w}=\rho (\nu)^2 V_{A}$ $\sim$ 6.43$\times$10$^{6}$ erg cm$^{-2}$ s$^{-1}$ with the $\nu$
taken as the maximum velocity amplitude ($\nu_{max}$). The estimated energy can balance typical
radiative and conductive losses of the loops and heat them to coronal temperature (Withbroe \& Noyes
1977; Ofman \& Wang 2008). However, the event reported here is a special event triggered by a flare
associated with filament activation. As a result, the mechanism in this event may not be responsible
for heating of all the coronal loops.

\begin{acknowledgements}
We thank the referee for his/her valuable suggestions.
The data are used courtesy of \emph{IRIS} and \emph{SDO} science teams. \emph{IRIS} is a NASA small
explorer mission developed and operated by LMSAL with mission operations executed at NASA Ames Research
center and major contributions to downlink communications funded by ESA and the Norwegian Space Centre.
This work is supported by the National Natural Science Foundations of China (11533008, 11790304, 11773039,
11673035, 11673034, and 11790300) and Key Programs of the Chinese Academy of Sciences (QYZDJ-SSW-SLH050).
\end{acknowledgements}

%
%

\clearpage

\end{document}